\def\la{\mathrel{\mathpalette\fun <}}
\def\ga{\mathrel{\mathpalette\fun >}}
\def\fun#1#2{\lower3.6pt\vbox{\baselineskip0pt\lineskip.9pt
\ialign{$\mathsurround=0pt#1\hfil##\hfil$\crcr#2\crcr\sim\crcr}}}
\relax
\documentstyle[12pt]{article}
\advance\textheight by \topskip
\begin{document}
\begin{flushright}
SU-ITP-92-18
\end{flushright}
\vskip .5cm
\begin{center}
{\Large\bf    PROGRESS IN THE THEORY OF THE }
\vskip 0.4cm
{\Large\bf ELECTROWEAK
PHASE TRANSITION}
\vskip 1.2cm
{\bf Andrei Linde} \footnote{E-mail: linde@physics.stanford.edu. On
leave from: Lebedev Physical Institute, Moscow }\\
Department of Physics, Stanford University, Stanford, CA 94305\\
\vskip 1  cm
{\large\bf ABSTRACT}
\end{center}
\begin{quote}
\ \ \ \ \ \ \ \ \ Recent progress in  the theory of the
electroweak phase transition is discussed.
It is shown,  that for the Higgs boson mass smaller than the
masses of W and Z  bosons, the phase transition is of the first
order. However, its strength is approximately 2/3 times less than
what follows from the one-loop approximation. This rules out
baryogenesis
in the minimal version of the electroweak theory with light Higgs
bosons.
The possibility of the strongly first order phase transition in

the theory
with superheavy Higgs bosons is considered.

\ \ \ \ \ We show that if the Yang-Mills field at high temperature
acquires a
magnetic mass $\sim g^2 T$, then the infrared problem  and
the problem of symmetry behavior at high
temperature effectively decouple from each other, no linear terms
appear in the effective potential in all orders of perturbation
theory and
 the symmetry in gauge theories at high
temperatures is actually restored. Even though the last statement was
never questioned by most of the authors, it was extremely difficult
to come to a reliable conclusion about it due to the
infrared problem in thermodynamics of non-Abelian gauge fields.

\ \ \ \ \ \ \ The phase transition occurs due to production and
expansion of critical bubbles.  A general analytic
expression for the probability of the bubble formation is obtained,
which may be used for study of tunneling in a wide class of theories.

\end{quote}
\vfill
\centerline {Invited Talk}
\centerline{Texas Symposium on Electroweak Baryon Number Violation}
\centerline{Yale University, March 1992}
\vfill

\pagebreak
\section{Introduction}

This talk is based on the results obtained in our  papers
with Michael Dine, Patrick Huet, Robert Leigh and Dmitri Linde
\cite{OurPaper}.
It contains also some more recent results on the infrared problem
in the electroweak theory,  on the existence of symmetry
restoration at a high temperature and on the possibility of the first
order phase transition in the theory with superheavy Higgs fields.

The existence of the phase transition in the electroweak theories was
discovered by David Kirzhnits twenty years ago
 \cite{[1]}.  A detailed theory of the phase
transition was proposed in 1974 by three groups of authors
independently (by Weinberg, Dolan and
Jackiw and by Kirzhnits and Linde  \cite{[1a]}), and soon the
theory of the electroweak phase transition became one of the
well established ingredients of modern cosmology.  Surprisingly
enough, the theory of this phase transition is still
incomplete.

In the first papers on this problem it was assumed that the phase
transition is of the second order  \cite{[1],[1a]}. Later
Kirzhnits and Linde showed  \cite{[1b]}
that in the gauge theories with many particles,
and especially with particles which are much more heavy than
the Higgs boson $\phi$, one should take into account corrections
to the high temperature approximation used in \cite{[1],[1a]}.
These corrections lead to the occurrence of cubic terms
$\sim g^3\phi^3 T$ in the expression for the effective potential
$V(\phi,T)$. As a result, at some temperature, $V$
acquires an extra minimum,
and the phase transition is first order  \cite{[1b]}.
Such phase transitions occur through the formation and subsequent
expansion of bubbles of the scalar field $\phi$ inside the
symmetric phase $\phi = 0$. A further investigation of this
question has shown that the phase transitions in grand unified
theories are always strongly first order  \cite{Linde81}. This
realization, as well as the mechanism of reheating of the universe
during the decay of the supercooled vacuum state suggested in
\cite{[1b],[11]}, played an important role in the development
of the first versions of the inflationary universe scenario
 \cite{Guth81}. (For a review of the theory of phase transitions
and inflationary cosmology see Ref.  \cite{[2]}.)

For a long time it did not seem likely that the
electroweak phase transition could have any dramatic consequences.
Even though the
possibility of a strong baryon number violation during the
electroweak phase transition was pointed out fifteen years ago
\cite{Tunn1,DS},
only after the  paper by Kuzmin, Rubakov and
Shaposhnikov \cite{KRS} was it realized that such processes do
actually occur and may erase all previously generated baryon
asymmetry of the universe.

Recently, the possibility that electroweak interactions may not only
erase but also produce the cosmic baryon asymmetry has led to
renewed interest in
the electroweak phase transition. A number of scenarios have been
proposed for generating the asymmetry  \cite{5} -- \cite{12}.
All of them require that the phase transition should be
strongly first order since otherwise the baryon asymmetry generated
during the phase transition subsequently disappears. In all of these
scenarios the asymmetry is produced near the walls of the bubbles of
the scalar field $\phi$.
Therefore it is necessary to make a much more thorough analysis
of the electroweak phase transition than the analysis which is
necessary for an approximate calculation of the critical temperature.

We will say that  the phase transition
is strongly first order if the ratio of the
Higgs field $\phi$ inside the bubble to the temperature $T$ is
larger than one, since otherwise the baryon asymmetry will be
washed out by nonperturbative effects.

This condition was used in  \cite{5,[8]} to impose a strong
constraint on the Higgs mass in the minimal version of the
electroweak theory, $m_H  \, \,  \la \,  \,  42$ GeV.  This, of
course, already contradicts the present experimental limits
$m_H  \, \, \,  \ga \,\, \, 57$
GeV  \cite{LEP}.  However,  more careful consideration of
various theoretical uncertainties indicated that the constraint
might be somewhat weaker, permitting $m_H$ up to $55$ GeV,
or possibly higher  \cite{[4]}.  In multi-Higgs
models  \cite{[8],8}, the limits are substantially weaker.

Before one can discuss details of the process of baryogenesis,
it is necessary to check that the results of our investigation of the
phase transition are reliable. This is not a trivial issue even in
the minimal electroweak theory. Indeed, as stressed in
Refs.  \cite{[1b],[11]}, each new order of perturbation
theory at finite temperature may bring a new factor of
$g^2 T/m \sim gT/\phi$ for the
theories with gauge boson masses $m \sim g\phi$.
This means that the results of the
one loop calculations may become unreliable at $\phi  \, \,  \la \,
\, gT$. Thus, it became very desirable to go beyond the one-loop
approximation.

An example of such an approach is given by the self-consistent
approximation elaborated in  \cite{[1b]}. In this approximation,
instead of the mass of a particle at zero temperature one uses its
temperature-dependent mass, taking into account the contribution from
the polarization operator. This method made it possible, in
particular,  to overcome unphysical difficulties related to imaginary
masses of scalar particles at small $\phi$.

Recently this approach was reinvented by many authors. Some of
the recent results obtained by this method were quite surprising.
For example, it was claimed that higher order corrections lead to the
appearance of a term in the effective potential $\sim - g^3\phi T^3$
 \cite{Hsu,Shap}. This term is linear in $\phi$; it is very large at
small $\phi$. Depending on its sign, it either  may remove the local
minimum of
$V(\phi,T)$ created by the cubic term $\sim - g^3\phi^3 T$, or it may
make this minimum much more deep.

Our investigation of this problem shows that if one is
careful with counting of Feynman diagrams, neither positive nor
negative linear
terms $\sim g^3 \phi T^3$ appear in the effective potential
  \cite{OurPaper}. Even though now the authors of Refs.
\cite{Hsu,Shap} agree that the linear terms $\sim g^3 \phi T^3$ are
absent, we will
repeat our main arguments here, since these arguments may allow us to
do much more than just say that there are no linear terms in order
$g^3$.
 A generalization of these arguments allows us to formulate the
conditions under which one can show,
 despite some uncertainties with higher order corrections,  that the
expectation value of the
scalar field $\phi$  at high temperatures actually disappears, $\phi
= 0$. Note, that this would be impossible in the presence of linear
terms of any magnitude and  sign.

However, higher order corrections do lead to a  significant
modification of the one-loop results. They lead to a
decrease of the cubic term $g^3 \phi^3 T$ by a factor $2/3$
\cite{OurPaper}.\footnote{A similar result was obtained also by
Carrington \cite{Carr}. However, her original results were different
from ours approximately by a factor of two. Consequently, they lead
to an impression that modification of the one-loop results should
lead to an {\it increase} of the strength of the first order phase
transition. At present, there is no disagreement between our
results.} This effect decreases the ratio $\phi/T$ at the point
of the phase transition by approximately the same factor $2/3$.
This makes baryogenesis virtually impossible in  the context of
the minimal standard model with $m_W \ga m_H \ga 57$ GeV.

All results discussed above were obtained in the context of the
theories with small coupling constants and light Higgs fields.
However, we do not really know whether the Higgs boson is light or
very heavy. If the Higgs boson is superheavy, $m_H \ga 10^3$ GeV, the
phase transition may become strongly first order. Even though this
possibility is extremely speculative, it may lead to important
consequences. Therefore we will discuss it  in  this paper.

Assuming that one knows the shape of the effective potential at small
$\phi$, one should still work hard to determine the ratio $\phi/T$ at
the point of the phase transition. One needs to know at what
temperature the transition actually occurs, and some details of
{\it how} it occurs.
At very high temperatures the effective potential of the Higgs
field, $V(\phi,T)$, has a unique minimum at the symmetric
point $\phi = 0$. As the temperature is lowered, a second minimum
appears. At a critical value $T_c$, this second minimum becomes
degenerate with the first one. However, the phase transition
actually occurs at a somewhat lower temperature, due to the
formation of bubbles of true vacuum which grow and fill the universe.
The usual way to study bubble formation is to use the euclidean
approach to tunneling at a finite temperature  \cite{[3]}. One should
find high-temperature  solutions, which describe the so-called
critical bubbles. Then one should calculate their action, which leads
to an exponential suppression of the probability of bubble formation.
Typically, these calculations are rather complicated, and
analytic results can only be obtained in a few cases.  One of
these is the thin wall approximation, which is valid (as in
the case of transitions at zero temperature) if the difference in
depth of the two minima of $V(\phi,T)$ is much smaller than the
height
of the barrier between them. In this case the radius of the bubble at
the moment of its formation is much larger than the size of the
bubble
wall, and the properties of the bubble can be obtained very easily.
However, the thin wall approximation in our case leads
to an underestimate of the
tunneling action by a factor of two.
 Fortunately, we were able to obtain a simple analytic
expression
which gives the value of the euclidean action for theories with an
effective potentials of a rather general type,
$V(\phi,T) = a \phi^2 - b\phi^3 + c\phi^4$.  We hope that this result
will be useful for a future investigation of
bubble formation in a wide class of gauge theories with
spontaneous symmetry breaking.

On the other hand, validity of the standard assumption that the
phase transition occurs due to formation of critical bubbles should
be verified as well. Kolb and Gleiser  \cite{[7]} and, more recently,
Tetradis  \cite{Tetr} have argued that the phase transition may occur
by a different mechanism, the formation of small (subcritical)
bubbles.
If this is the case, the transition is completed earlier and by a
different mechanism than in the conventional picture.  While this
idea
is very interesting, we will argue (see also \cite{Kaj} and the talk
of Anderson at this Conference)
that it is only relevant in cases
where the transition is very weakly first order and the euclidean
action corresponding to critical bubbles is not much larger than one.
This is not the case for the strongly first order phase transitions,
where the relevant value of the euclidean action at the moment of
the transition is $S \sim 130 - 140$.

\section{The Phase Transition}

Let us consider the form of the effective
potential at finite temperature. Contributions of particles of
a mass $m$ to $V(\phi,T)$ are proportional to $m^2\,T^2$, $m^3
\,T$ and
$m^4 \ln (m/T)$. We will assume that the Higgs boson mass
is smaller than the masses of W and Z bosons and the
top quark, $m_H < m_W, m_Z, m_t$. Therefore we will
neglect the Higgs boson contribution to $V(\phi,T)$.

The zero temperature potential, taking into account
one-loop corrections, is given by  \cite{[2]}
\begin{equation}\label{3}
V_0 = - {\mu^2\over 2}\phi^2 + {\lambda\over 4} \phi^4 +
2Bv_o^2\phi^2 - {3\over 2} B\phi^4 + B \phi^4 \ln({\phi^2\over
v_o^2}) \ .
\end{equation}
Here
\begin{equation}\label{4}
B = {3\over 64 \pi^2 v_o^4} (2 m_W^4 +
m_Z^4 - 4 m_t^4) \ ,
\end{equation}
$v_o = 246$ GeV is the value of the scalar field at the minimum
of $V_0$, $\lambda = \mu^2/v_o^2$,
$m^2_H = 2\mu^2$. Note that these relations between $\lambda, \mu,
v_o$ and the
Higgs boson mass $m_H$, which are true at the classical level,  are
satisfied even
with an account taken of the one-loop corrections. This is an
advantage of the normalization conditions used in  \cite{[2]}.
An expression used in  \cite{[5]} is equivalent to this expression up
to an obvious change of variables.

At a finite temperature, one should add to this expression the term
\begin{equation}\label{5}
V_T = {T^4\over 2 \pi^2} \Bigl(6I_{-}(y_W) + 3I_{-}(y_Z) -
6I_{+}(y_t)\Bigr) \ ,
\end{equation}
where $y_i = M_i\phi/v_o T$, and
\begin{equation}\label{6}
I_{\mp}(y) = \pm \int_{0}^{\infty} dx
\ x^2 \ln (1  \mp e^{- \sqrt{x^2+y^2}}) \ .
\end{equation}
The results of our work are based on numerical calculation of these
integrals, without making any specific approximations  \cite{[4]}.
However, in the large temperature limit it is sufficient to use an
approximate expression for $V(\phi,T)$  \cite{[1],[5]},
\begin{equation}\label{7}
V(\phi,T) = D (T^2 - T_o^2) \phi^2 - E T \phi^3 +
{\lambda_T\over 4} \phi^4 \ .
\end{equation}
Here
\begin{equation}\label{8}
D = {1\over 8v_o^2} ( 2 m_W^2 +
m_Z^2 + 2 m_t^2) \ ,
\end{equation}
\begin{equation}\label{9}
E =  {1\over 4\pi v_o^3} ( 2 m_W^3 +
m_Z^3) \sim 10^{-2} \ ,
\end{equation}
\begin{equation}\label{10}
T^2_o = {1\over 2D}(\mu^2 - 4Bv_o^2) =
{1\over 4D}(m_H^2 - 8Bv_o^2) \ ,
\end{equation}
\begin{equation}\label{11}
\lambda_T = \lambda - {3\over 16 \pi^2 v_o^4}
\left( 2 m_W^4 \ln{m^2_W\over a_B T^2} +
m_Z^4 \ln{m^2_Z\over a_B T^2} -
4 m_t^4 \ln{m^2_t\over a_F T^2}\right) \ ,
\end{equation}
where $\ln a_B = 2 \ln 4\pi - 2\gamma \simeq 3.91$,
$\ln  a_F = 2 \ln \pi - 2\gamma \simeq 1.14$.

It will be useful for our future discussion to identify
several `critical points' in the evolution of $V(\phi,T)$.

At very high temperatures the only minimum of $V(\phi,T)$ is
at $\phi = 0$. A second minimum appears at $T= T_1$, where
\begin{equation}\label{12}
T^2_1 = {T^2_o \over {1 - 9 E^2/8\lambda_{T_1}D}} \ .
\end{equation}
The value of the field $\phi$ in this minimum at $T = T_1$ is equal
to
\begin{equation}\label{13}
\phi_1 = {3ET_1 \over 2\lambda_{T_1}} \ .
\end{equation}
The values of  $V(\phi,T)$ in the two minima become equal
to each other at the temperature $T_c$, where
\begin{equation}\label{14}
T^2_c = {T^2_o \over {1 -  E^2/\lambda_{T_c}D}} \ .
\end{equation}
At that moment the field $\phi$ in the second minimum
becomes equal to
\begin{equation}\label{15}
\phi_c =  {2ET_c \over \lambda_{T_c}}\ .
\end{equation}
The minimum of $V(\phi,T)$ at $\phi = 0$ disappears at the
temperature $T_o$, when the field $\phi$ in the second
minimum becomes equal to
\begin{equation}\label{16}
\phi_o = {3ET_o \over \lambda_{T_o}}\ .
\end{equation}

\section{Infrared Problems and Reliability of the Perturbation
Expansion}

In our previous discussion, we have considered only the one loop
corrections to the effective potential.  In this section we
discuss the role of higher order corrections.

It is well known that, in field theories of massless particles,
perturbation theory at finite temperature
is subject to severe infrared divergence problems.  For small
values of the scalar field, the gauge bosons (and near the
phase transition, the Higgs boson) are nearly massless;
as a result, as was pointed out in the early work on this subject
\cite{[1b],[11]}, one cannot reliably compute the effective
potential for very small $\phi$.  The problem is that the higher
order
corrections in coupling constants may contain terms of
the type of $\Bigl({g^2\, T\over m}\Bigr)^N$.
As a result, higher order corrections go out of control for
$m < g^2 T$. For scalar particles this happens near the
critical point only. Indeed,  scalar particles have
masses $m \sim gT \gg g^2T$ in the high temperature limit.
However, if one takes into account gauge invariance, it can
be shown that ``magnetic'' components of vector particles
cannot acquire any contribution to their ``masses'' larger
 than $g^2 T$.

At this point one should be more precise. The Green function
of the vector field is singular at
$k^2 \sim g^2 T^2$. In this sense one may speak about the
vector field mass $\sim gT$. However, the Green function
of the vector field at a finite temperature does not have a
simple pole singularity. For example, in addition to the
singularity at $k^2 \sim g^2 T^2$, the Green function of a
photon has a singularity at $k_0 = 0, \vec{k} \rightarrow 0$.
It is this singularity that is responsible for all infrared problems
in quantum statistics of gauge fields, since the Green functions at
$k_0 = 0$ give the leading infrared divergent contribution to
thermodynamical sums \cite{[11],L,LGPY}.
Investigation of the infrared problem in gauge theories without
spontaneous symmetry breaking has shown that the ``magnetic mass''
(corresponding to the limit  $k_0 = 0, \vec{k} \rightarrow 0$) may
appear in the non-Abelian theories, but it cannot be larger than
$O(g^2) T$.

Thus, in the absence of spontaneous symmetry breaking, or at
$\phi \la gT$, when the magnetic mass of the vector particles
become smaller than $g^2T$, perturbative results may
become unreliable. One may wonder, therefore, is it possible
that some unusual contribution to the effective potential at
$\phi \la gT$ may alter our results.

Recently, in a very interesting paper, Brahm and Hsu  found that
at small $\phi$, higher order corrections to the scalar
field contribution to the effective potential may produce a
large negative linear term $-\, g^3\phi T^3$, which eliminated
any trace of a first order transition.

On the other hand, Shaposhnikov considered higher order
corrections to the vector particle contribution to $V(\phi,T)$
and found a large positive term $+\, g^3\phi T^3$which made
the phase transition  strongly first order
($\phi/T\, >\, 1$) even for $m_H \sim 64$ GeV \cite{Shap}.

We will show that neither positive nor negative
linear terms appear  in the expression for
$V(\phi,T)$ if one studies higher order corrections paying
particular attention to the correct counting of Feynman diagrams.
\cite{OurPaper}. (Additional information on this problem is
contained in the talks by Michael Dine and Robert Leigh at
 this conference.)

We will consider here for simplicity the contribution of the scalar
particles and the W bosons
only; adding the contribution of Z bosons is trivial.
As we have already noted, for questions of infrared behavior,
fermions may be ignored. Coulomb gauge,
$\vec{\nabla}\cdot\vec{W }= 0$, is particularly convenient for the
analysis,
though the problem can be analyzed in other gauges as well. In this
gauge, the vector field propagator
$D_{\mu\nu}$ after symmetry breaking (and after a proper
diagonalization) splits into two pieces, the
Coulomb piece, $D_{00}$, and the transverse piece, $D_{ij}$.  For
non-zero values of the discrete frequency, $\omega_n=2\pi n T$,
the Coulomb piece mixes with the 'Goldstone' boson. However,
for the infrared problems which concern us here, we are only
interested in the propagators at zero frequency.  For
these there is no mixing.  One has \cite{[1b]}
\begin{equation}\label{i1}
D_{00}(\omega=0,\vec{k}) = {1 \over \vec{k}^2 + m^2_W(\phi)}
\end{equation}
and
\begin{equation}\label{i2}
D_{ij}(\omega=0,\vec{k}) = {1 \over\vec{k}^2 +
m^2_W(\phi)}P_{ij}(\vec{k}) \ ,
\end{equation}
where $P_{ij}=\delta_{ij} -{k_i k_j \over \vec{k}^2}$. The mass of
the
vector field $W$ at the classical level is given by $m_W = g v_o /2$.
Propagators of the Higgs field $\phi$ and of the 'Goldstone' field
$\chi$ in this gauge are given by
\begin{equation}\label{i3}
D_\phi(\vec{k}) = {1 \over  \vec{k}^2 + m^2_\phi} \ ,
\end{equation}
\begin{equation}\label{i4}
D_\chi(\vec{k}) = {1 \over \vec{k}^2} \ .
\end{equation}

Let us review several ways of obtaining the standard one-loop
expression for the cubic term in the effective potential, eq.
(\ref{7}). The most straightforward
is to carefully expand eq. (\ref{5}) for the effective potential in
$y_W = {m_W\over v_o T} = {g \phi\over 2T}$. Indeed, the contribution
of $W$-bosons to the effective potential at $T > m_W(\phi)$ is given
by
\begin{eqnarray}\label{i7}
V_W(\phi,T) &=& 2\times 3\times \Bigl( - {\pi^2\over 90} T^4 + {
m^2_W(\phi)\over 24} T^2 - {m^3_W(\phi) \over 12 \pi} T
+ \cdots\Bigr) \nonumber\\
  &=& 2\times 3\times \Bigl( - {\pi^2\over 90} T^4 + {
g^2 \phi^2\over 96} T^2 - {g^3\phi^3 \over 96\pi} T
+ \cdots\Bigr) \ .
\end{eqnarray}
Here the expression in brackets coincides with the contribution of a
scalar field
with mass $m_W$; the factor $2$ appears since there are two
$W$-bosons with opposite charges, while
the factor 3, which will be particularly important
in what follows, corresponds to the two transverse and one
longitudinal degrees of freedom with mass $m_W$.

Alternatively, we can obtain the cubic term by looking
directly at the one-loop Feynman diagrams.  For this purpose,
it is only necessary to examine the zero frequency contributions.
Certain diagrams containing {\it four} external lines of the
classical scalar field naively give a contribution proportional to
$g^4 \phi^4$; the cubic term arises because the zero
frequency integrals diverge for small mass as $T / m_W
\sim T / g \phi$.

Consider, in particular, the zero frequency part of the expression
for the one loop free energy in momentum space.  It is simplest
to compute the tadpole diagrams for $dV / d\phi$ and afterwards
integrate with respect to $\phi$.
The transverse gauge bosons give a contribution
\begin{equation}\label{trzero}
{dV_{{\rm tr}}\over d\phi} =
 2\times  {g^2 \phi T \over 2}\ \int {d^3 k \over (2 \pi)^3}
 {1\over \vec{k}^2 + m_W^2}
   = -2\times {g^2\phi\; T\over 8\pi}\, \sqrt{m_W^2} \ ,
\end{equation}
where, by keeping only the zero frequency mode, we have
dropped terms which are analytic in $m^2$. The Coulomb lines give
half the result of
eq. (\ref{trzero}). Integration of the total vector field
contribution correctly represents the cubic term in (\ref{i7}).

A complete gauge boson contribution to the tadpole, including
  the non-zero frequency modes, is \cite{[1b]}
\begin{equation}\label{i9}
{dV_W(\phi,T)\over d\phi}  = 2\times 3\times  {g^2\phi\over 48} \
\Bigl( T^2 -
{3 m_W T \over\pi }+ \cdots\Bigr) = 2\times 3\times  {g^2\phi\over
48} \ \Bigl( T^2 - {3\,g\phi T \over 2 \pi }+ \cdots\Bigr) \ .
\end{equation}
One can easily check that integration of this expression with respect
to $\phi$ gives eq. (\ref{i7}).

With these techniques, we are in a good position to study higher
order corrections to the potential.
The authors of Refs. \cite{Hsu,Shap} found a linear
contribution to the potential by substituting the mass found at one
loop back into the one loop calculation.  The effective
masses-squared of both scalar particles and of the Coulomb field
contain terms of the form $\sim g^3 T \phi$, which, upon substitution
in (34), give linear terms. But this procedure  is not always
correct. It is well known that the sum of the geometric progression,
which appears after the insertion of an arbitrary number of
polarization operators $\Pi(\phi,T)$ into the propagator
$ (k^2 + m^2)^{-1}$, simply gives $(k^2 + m^2 + \Pi(\phi,T))^{-1}$.
Therefore one can actually use propagators
$(k^2 + m^2 + \Pi(T))^{-1}$, which contain the effective mass-squared
$m^2 + \Pi(\phi,T)$ instead of $m^2$. However, this trick
with the geometric progression does
not work for the closed loop diagram for the effective potential,
which contains $\ln (k^2 + m^2)$. A naive substitution of the
effective mass squared $m^2 + \Pi(\phi,T)$ instead of $m^2$ into
$\ln (k^2 + m^2)$ corresponds to a wrong counting of
higher order corrections.

This does not mean that there is no regular way to make this trick
for the effective potential. One may add and subtract from the
Lagrangian the term $-{1\over 2}\, \phi^2\, \Pi(\phi(T),T)$ to the
Lagrangian,
and a similar term for the vector field as well. Here
$\Pi(\phi(T),T)$ is the polarization operator with an account taken
of all daisy and superdaisy diagrams,  $\phi(T)$ is a classical
field, not an operator.  Then the effective mass (at zero momentum)
becomes renormalized, $m^2 \rightarrow m^2 + \Pi(\phi,T)$, but one
should add some extra diagrams containing the insertion
$-{1\over 2}\, \phi^2
\, \Pi(\phi(T),T)$. These diagrams were not considered in
\cite{Hsu,Shap}. Here one can clearly understand the difference
between calculating effective potential and tadpoles. If one uses
this method to calculate  the tadpole diagrams,  insertions of
$-{1\over 2}\,
\phi^2 \, \Pi(\phi(T),T)$ self-consistently cancell the diagrams
which
would generate extraneous polarization operator corrections to
already corrected effective mass $m^2(\phi(T),T) = m^2 +
\Pi(\phi,T)$. This gives us the standard prescription of simply
substituting $m^2(\phi(T),T) = m^2 + \Pi(\phi,T)$ instead of $m^2$ in
all propagators $ (k^2 + m^2)^{-1}$.

Thus, the simplest  way to take into account high temperature
corrections to masses of vector and scalar particles without any
problems with combinatorics is to compute tadpole diagrams for
${d\,V\over d\phi}$; these are then trivially integrated to give the
potential. One can easily check by this method  that  no linear terms
appear
in the expression for  $V(\phi,T)$. Indeed, at a given temperature
and effective mass, the tadpoles are linear in $\phi$ (see e.g.
equation (\ref{i9})). To take into account the mass renormalization
in the tadpoles, one
 should substitute the effective mass squared $m^2 + \Pi(\phi,T)$
into the one-loop expression for the tadpole
contribution; as we explained above (see also (\cite{[1b]})), this is
a correct and unambiguous procedure for tadpoles.
\begin{equation}\label{trzero1}
{dV_{{\rm tr}}\over d\phi} =
 2\times  {g^2 \phi T \over 2}\ \int {d^3 k \over (2 \pi)^3}\
 {1\over \vec{k}^2 + m^2+ \Pi(\phi,k,T)} \ .
\end{equation}
This expression could lead to a linear term in the effective
potential only if the integral would behave as $\phi^{-1}$ in at
small $\phi \rightarrow 0$. However, in our case this is impossible,
since the polarization operator calculated in \cite{Hsu,Shap}  at
small $k$ and $\phi$ is {\it nonnegative}, and the integral converges
to a constant.  Therefore its subsequent integration with respect
to $\phi$, which gives the
correction to the effective potential,  is quadratic  in $\phi$,
{\it i.e.} it does not contain any linear terms.

We should emphasize that our approach effectively takes into account
{\it all} diagrams considered in \cite{Hsu}, \cite{Shap}, but with
correct combinatorics. Note also, that this approximation works even
if the polarization operator depends on the classical field
$\phi$.\footnote{We disagree with the recent claim made in
 \cite{EQZ} that our approximation works only if the polarization
operator $\Pi$ does
not depend on $\phi$, and that we do not take into account
`subleading' diagrams which  give rise
to dangerous linear terms obtained in \cite{Hsu}, \cite{Shap}.
Indeed, in \cite{OurPaper} we did not take into account diagrams of
the type  shown in Fig. 2; see their discussion below. However, these
diagrams were not considered in  \cite{Hsu}, \cite{Shap} as well,
since they do not lead to the linear terms in the order
$g^3$.} An apparent
triviality of the investigation with the help of tadpoles is not due
to its incompleteness, but due to the power and simplicity of this
method,  which was elaborated in \cite{[1b]}.
In particular, it was even unnecessary for us to use a particular
expression for the polarization operator, as far as it is
nonnegative. We will return to this issue shortly.

Even though there are no linear terms $\sim g^3 \phi T^3$, higher
order corrections do have a dramatic effect
on the phase transition. This effect is a modification of the cubic
term.

As we have shown above, the cubic term appears due to the
contribution of zero modes, $\omega_n = 2\pi nT = 0$. This makes it
particularly easy to study its modification by high order effects.
Indeed, it is well known that the Coulomb field at zero frequency
acquires the Debye `mass',  $m^2_D = \Pi_{00}(\omega_n = 0, \vec{k}
\to 0) \sim g^2 \, T^2$.
This leads to an important modification of the Coulomb
propagator (\ref{i1}):
{\begin{equation}\label{coulombloop}
D_{00}(\vec{k}\to 0)={1 \over \vec{k}^2 + m_D^2 + m_W^2(\phi)} \ .
\end{equation}
For the values of $\phi$ of interest to us, $m_D^2 \gg m_W^2(\phi)$.
Thus, repeating the calculation of the cubic term,
the Coulomb contribution disappears.  However, the transverse
contribution, which is two times larger than the Coulomb one,
is unaffected at this order, due to the vanishing of the `magnetic
mass'  \cite{[11],L,LGPY}.  As a
result, the  cubic term does not disappear,  but it is diminished by
a factor\footnote{There was also a claim that the cubic term
disappears completely \cite{evans}, but recently this claim was
withdrawn.} of $2/3$:
\begin{equation}\label{newE}
E =  {1\over 6\pi v_o^3} ( 2 m_W^3 +
m_Z^3) \ .
\end{equation}

 This small correction proves to be very significant. Indeed,
eqs. (\ref{15}), (\ref{16}) show that the ratio of the scalar field
$\phi$ to the
temperature at the moment of the phase transition is proportional to
$E$, i.e.
to the cubic term. Actually, the dependence is even slightly
stronger, since
for smaller  E the tunneling occurs earlier.  Even before
the reduction of the cubic term  was taken into account, the ratio
$\phi/T$ for $m_H \, \, \ga \, \, 57$ GeV was slightly less than
the critical value $\phi/T \approx 1$. The
decrease of this quantity by a factor of  $2/3$ makes it absolutely
impossible to preserve the
baryon asymmetry generated during the phase transition in the minimal
model of electroweak interactions with $m_H \, \,  \ga \, \, 57$ GeV.

Are these results completely reliable? The effective coupling
constant of interactions
between W bosons and Higgs particles is $g/2$. In this case, a
 general investigation of the infrared
problem in the non-Abelian gauge theories at a finite temperature
suggests that  the results which
we obtained are reliable for
$\phi \  \,  \ga \,  \ {g\over 2} \ T \sim T/3 $ \cite{[11]},
\cite{L,LGPY}.
Thus,  a more detailed investigation is needed to study behavior
of the theories  with $m_H \, \,  \ga \, \,  10^2$ GeV near the
critical temperature, since the scalar field,
which appears at the moment of the phase transition
in these theories, is very small. However, we expect
that our results are reliable for strongly first order phase
transitions with $\phi\,\, \,  \ga \, \,\, T/3$, which is quite
sufficient
to study (or to rule out) baryogenesis in the electroweak theory.
Recent investigation of higher order corrections to this theory
indicates \cite{BHG} that these results may be reliable even down to
$\phi\,\,  \sim  \,\, gT/10$.

Finally, we would like to address a fundamental question:  since
the theory for $\phi\ll gT$ is infrared divergent, can we definitely
establish that the symmetry is restored at high temperature,
or is it possible that $\phi$ always has some small, non-zero
value?  Indeed, if our approximation breaks down at
$\phi\,\,  \la  \,\, gT$, how do we know that the symmetry
restoration actually takes place, i.e. $\phi = 0$ at $T > T_o$?

To address this question, we can work far away from the
critical point, at $T- T_o \gg T_o$.
The best approach, as before, is to study all possible higher order
tadpole diagrams, see Figs. 1, 2.

There are two different classes of diagrams to be considered.
External line of the scalar field may split either into two lines of
the vector field, Fig. 1, or into two lines of vector field and one
line of scalar field, Fig. 2. All diagrams of the first type can be
represented  as the trivial one-loop diagram, plus the  diagrams with
an arbitrary number of polarization operator insertions. The simplest
diagram of that type is shown in Fig. 1. The black circle  stands for
the exact polarization operator $\Pi(\phi,k,T)$, including all higher
order corrections. The sum of all these diagrams gives us the
one-loop diagram with an exact Green function of the vector field
instead of the free field propagator. In other words, as usual, one
must add polarization operator to the mass squared of the vector
field.

The behavior of $\Pi(\phi,k,T)$ at $k > g^2 T$ is known
perturbatively. It leads only to high-order corrections to
$V(\phi,T)$, which  do not contain any nonanalytic terms, such as the
dangerous  linear terms in $\phi$. The only possible
source of problems is our absence of knowledge of $\Pi(\phi,k,T)$ at
$\phi \la gT$, $k_0 = 0$,  $|\vec{k}|\la g^2 T$.
 Indeed, in this domain all higher order corrections to
$\Pi(\phi,k,T)$ are equally important. However, the
consequences of this uncertainty may appear  not very significant.

 Indeed,  consider again the most dangerous part of the tadpole
diagram, eq. (\ref{trzero}), and add $\Pi(\phi,k,T)$ to
$m^2_W(\phi)$:
 \begin{equation}\label{trzero2}
{dV_{{\rm tr}}\over d\phi} =
 2\times  {g^2 \phi T \over 2}\ \int {d^3 k \over (2 \pi)^3}\
 {1\over \vec{k}^2 + {g^2 \phi^2\over 4}+ \Pi(\phi,k,T)} \ .
\end{equation}
The part of the integral in the domain of uncertainty, $|\vec{k}|\la
g^2 T$, is given by
 \begin{equation}\label{trzero3}
 {g^2 \phi T \over 2\pi^2}\ \int_{0}^{g^2T} {k^2 d k \over k^2 + {g^2
\phi^2\over 4} + \Pi(\phi,k,T)} \ .
\end{equation}
On dimensional grounds, one expects that at  $g\phi \ll g^2 T$ and
$|\vec{k}| \ll g^2 T$ the polarization operator has some value of the
order  $g^4 T^2$
\cite{[11],L,LGPY}. This follows from the fact that the most infrared
divergent part of the theory corresponds to the three-dimensional
theory, with $g^2 T$ being the only mass (or coupling constant)
scale.

One cannot exclude a possibility  that the polarization operator   is
proportional to $-g^2 |\vec{k}| T$, or it is of the order  $g^4 T^2$,
but is negative,
 and the integral in (\ref{trzero3}) diverges in the limit $\phi
\rightarrow 0$.
  One should keep this
possibility in mind, since it may lead to interesting and unusual
consequences, like Bose-condensation or even crystallization of the
Yang-Mills field at high temperature \cite{L}. Indeed, the  tadpole
integral has a simple interpretation in terms of integration over the
occupation numbers of bose fields. Large contribution to this
integral may be interpreted as a result of Bose condensation of
particles in a state with a nonvanishing momentum. In our case, this
effect may also lead to absence of a complete symmetry restoration at
$T > T_o$. Note, that this effect may occur only at $\phi \la gT$,
as we anticipated. However, it is not quite clear that this
possibility is
physically viable.  Whereas occupation numbers may be large, they can
hardly be negative.

The standard (and most conservative) assumption is that $0 \leq
\Pi(\phi =
0, k = 0) \la g^4T^2$ \cite{[11],L,LGPY}. This corresponds to
generation of a magnetic mass $0 \leq m^2 \la g^4 T^2$. In such case
we may
estimate the corresponding integral at small $\phi$ as
  \begin{equation}\label{trzero4}
 {g^2 \phi T \over 2\pi^2}\ \int_{0}^{g^2T} {k^2 d k \over k^2 +
O(1)\, g^4 T^2} \sim g^4\phi T^2 \ .
\end{equation}
This term after integration over $\phi$ does not give any linear
terms in $\phi$. It gives just a small
correction to the quadratic part of the effective potential, $\Delta
V \sim g^4 \phi^2 T^2$. Such corrections  do not alter our
conclusions
concerning symmetry restoration at high temperature. Note, that this
term corresponds to the sum of {\it all}\, most dangerous
contributions to the tadpole diagrams of the type of Fig.1, to all
orders in $g^2$.

Now let us consider the diagrams which contain internal lines of
scalar field,  Fig. 2. These diagrams may be dangerous near the
critical point, where the scalar fields are almost massless, but
they are much less dangerous than the diagrams of
the first class at the temperature much higher than critical. The
reason is that at high temperature the scalar field acquires a large
mass $m^2 \sim g^2 T^2 \gg g^4 T^2$. Therefore scalar particles by
themselves do not  lead to any infrared problems outside of  a small
vicinity of the critical point. The presence of such heavy particles
effectively cuts infrared divergencies in the diagrams with vector
particles as well. One can easily check that the diagram with one
vector loop, Fig 2a, gives the contribution $g^4\phi^2T^2$ to the
effective potential, the diagram with two vector loops gives
$g^5\phi^2T^2$, the diagram with three vector loops gives
$g^6\phi^2T^2$. Starting with this diagrams, infrared problem becomes
manifest in that each diagram of this type with higher number of
vector loops gives the contribution of the same order $g^6\phi^2T^2$.

Thus, the infrared problem in thermodynamics of gauge fields does not
permit us to calculate the effective potential at high temperature to
all orders of perturbation theory. The diagrams Fig. 2 contain
uncertainties at the level of $g^6\phi^2T^2$; the diagrams Fig. 1
contain uncertainties at the level of $g^4\phi^2T^2$. However,
neither of these diagrams produce linear terms in $\phi$, unless
the Green functions of a massless Yang-Mills field has a
pathological behavior at large temperature. As for the
quadratic terms, they can be calculated at least with an accuracy up
to $g^3\phi^2T^2$, or maybe even up to $g^4 \ln g\ \phi^2T^2$. This
is
quite sufficient to calculate the critical temperature and to make a
conclusion that at the temperature higher than critical the scalar
field $\phi$ vanishes.

These considerations indicate that
the situation with the phase transitions in the non-Abelian
gauge theories is probably  the same as in the standard
case: infrared problems may prevent a simple description of the
phase transition in a small vicinity of the
critical point (unless the phase transition is strongly first order),
but everywhere outside this region, the symmetry behavior
of gauge theories can be described in a reliable way.

\section{First order phase transitions with superheavy Higgs?}

In our previous investigation we neglected the contribution of Higgs
bosons to  the one-loop effective potential.  The reason  was very
simple: We have seen that the increase of the Higgs boson mass
decreases the strength of the phase transition. This can be easily
understood by considering a model of a single scalar field with the
effective potential
\begin{equation}\label{u}
V_0 = - {\mu^2\over 2}\phi^2 + {\lambda\over 4} \phi^4 \equiv -
{m_H^2\over 4}\phi^2 + {\lambda\over 4} \phi^4  \ .
\end{equation}
Near the point of the phase transition, where the high-temperature
approximation works well, one may investigate the symmetry behavior
in this theory in a self-consistent approximation suggested in
\cite{[1b]}, where only cactus diagrams should be evaluated. In this
approximation the effective mass of the scalar field and the first
derivative of the effective potential in its extremum are simply
related to each other:
\begin{equation}\label{u1}
m^2(T,\phi) =  3\lambda \phi^2  - \mu^2 + \Pi(T,m(T,\phi)) \ ,
\end{equation}
and
\begin{equation}\label{u2}
{1\over \phi} {d V^1\over d \phi} = \Pi(T,m(T,\phi)) \ .
\end{equation}
Here $V^1$ is the one-loop contribution to the effective potential,
\begin{equation}\label{u3}
{1\over \phi} {d V^1\over d \phi} = {3\lambda\over
2\pi^2}\int_0^\infty {k^2 \,dk
\over \sqrt {k^2 + m^2(T,\phi)}\left(\exp{\sqrt {k^2 +
m^2(T,\phi)}\over T} - 1\right)} \ .
\end{equation}
At the minimum of the effective potential with $\phi(T) \not = 0$
\begin{equation}\label{u4}
{d V^1\over d \phi} +  \lambda \phi^3  - \mu^2\phi  = 0 \ ,
\end{equation}
which, together with (\ref{u1}) and (\ref{u2}),  gives $m^2 =
2\lambda \phi^2(T)$ and
\begin{equation}\label{u5}
{1\over \phi} {d V^1\over d \phi} = {3\lambda\over
2\pi^2}\int_0^\infty {k^2 \,dk
\over \sqrt {k^2 + 2\lambda\phi^2(T)}\left(\exp{\sqrt {k^2 +
2\lambda\phi^2(T)}\over T} - 1\right)} = {\lambda T^2\over 4}\left(1
- {3\sqrt{2\lambda}\phi(T)\over \pi T} + ... \right).
\end{equation}
On the other hand, the local minimum of $V(\phi)$ at $\phi = 0$
disappears when $m(T,0) = 0$. This gives  the critical temperature
\begin{equation}\label{u6}
T_o = 2 v \ ,
\end{equation}
where $v = \mu/\sqrt\lambda = 246$ GeV, $v > \phi(T)$. For
small $\lambda$ the last term in eq. (\ref{u5})  is small at $T
\approx T_o$, which
implies that the phase transition is  weakly first order. Actually,
we cannot even say from eq. (\ref{u5}) whether  the phase transition
is second order or weakly first order, since near the critical point
the higher order corrections are large \cite{[1b]}.

Let us nevertheless use eq. (\ref{u5}) to make a bold estimate of
conditions under which
the phase transitions could be strongly first order. From eqs.
(\ref{u5}),  (\ref{u6}) it follows that the jump of the scalar field
at the critical temperature
(assuming that $T_o$ is close to the temperature of the phase
transition) is given by ${3\sqrt{2\lambda}T\over 4\pi }$. It is
larger than the temperature T (which is the condition for
baryogenesis) if $\lambda \ga 9$, or, equivalently,
\begin{equation}\label{u7}
 m_H \ga 10^3 \ GeV \ .
\end{equation}
For obvious reasons, this estimate should not be
taken as a serious indication of existence of  strongly first order
phase
transitions and baryogenesis with superheavy Higgs bosons.
However,  the stakes are high, and
the possibility to have a strongly first order phase transition and
baryogenesis in the strong coupling regime with superheavy Higgs
bosons (technicolour?) should not be overlooked.

\section{Bubble Formation}

In the previous section we noted that the two minima of
$V(\phi,T)$ become of the same depth at the temperature
$T_c$, eq. (\ref{14}). However, tunneling with formation of bubbles
of the field $\phi$ corresponding to the second minimum starts
somewhat later, and it goes sufficiently fast to fill the whole
universe with the bubbles of the new phase only at some
lower temperature T when the corresponding euclidean action
suppressing the tunneling becomes less than 130 -- 140
\cite{7,[4],[5]}.  In \cite{OurPaper}
(see also  \cite{[4]}) we performed a numerical
study of the probability of tunneling. Before reporting our results,
we will remind the reader of some basic concepts of the theory of
tunneling at a finite temperature.

In the euclidean approach to tunneling (at zero temperature)
\cite{[6]}, the probability of bubble formation in quantum field
theory is proportional to $\exp (-S_4)$, where $S_4$ is the
four-dimensional Euclidean action corresponding to the
tunneling trajectory. In other words, $S_4$ is the instanton action,
where the instanton is the solution of the euclidean field equations
describing tunneling. A generalization of this method for tunneling
at a very high temperature  \cite{[3]} gives the probability of
tunneling per unit time per unit volume
\begin{equation}\label{19}
P \sim A(T) \cdot \exp(- {S_3\over T}) \ .
\end{equation}
Here $A(T)$ is some subexponential factor roughly
of order $T^4$;  $S_3$ is a three-dimensional instanton action.
It has the same meaning (and value) as the fluctuation of the free
energy $F = V(\phi(\vec x),T)$ which is necessary for bubble
formation. To find $S_3$, one should first find an $O(3)$-symmetric
 solution, $\phi(r)$, of the  equation
\begin{equation}\label{20}
{d^2\phi\over dr^2} + {2\over r}\,{d\phi\over dr} = V^\prime(\phi) \
, \end{equation}
with the boundary conditions $\phi(r=\infty) = 0$ and
$d\phi/dr|_{r=0} = 0$. Here $r = \sqrt {x^2_i}$; the $x_i$ are the
euclidean coordinates, i = 1,2,3. Then one should calculate the
corresponding action
\begin{equation}\label{21}
S_3 = 4\pi \int_{0}^{\infty} r^2 \ dr\bigl[{1\over 2}
\left({d\phi\over dr}\right)^2 + V(\phi(r),T)\bigr] \ .
\end{equation}

Usually it is impossible to find an exact solution of eq. (\ref{20})
and to calculate $S_3$ without the help of a computer. A few
exceptions to this rule are given in Refs.  \cite{[2],[3]}.
One of these exceptional cases is
realized if the effective potential has two almost degenerate minima,
such that the difference $\varepsilon$ between the values of
$V(\phi,T)$ at these minima is much smaller than the energy barrier
between them. In such a case the thickness of the bubble wall at the
moment of its formation is much smaller than the radius of the
bubble, and the action $S_3$ can be calculated exactly as a
function of the bubble radius $r$, the energy difference
$\Delta V$ and the bubble wall surface energy $S_1$:
\begin{equation}\label{22}
S_3 = - {4\pi\over 3} r^3 \Delta V + 4\pi r^2 S_1 \ ,
\end{equation}
where
\begin{equation}\label{23}
S_1 = \int_0^\infty d\phi \sqrt{2V(\phi,T)} \ .
\end{equation}
The radius of the critical bubble $r$ can be found by finding an
extremum of $S_3(r)$.  However, one must be very careful when
using these results.
Indeed, as can be easily checked, this extremum is not
a {\it minimum} of the action, it is a {\it maximum}.
Therefore, the action corresponding to the true solution of eq.
(\ref{20}) will
be higher than the action of any approximate solution. As a result,
one can strongly overestimate the tunneling probability by
calculating it outside the limit of validity of the thin wall
approximation. In our case the thin wall approximation underestimates
the tunneling action by a factor of two, i.e. it gives the
probability of tunneling about $e^{-100}$ where the correct answer is
$e^{-200}$.  If the only thing one wishes to know is the time when
the tunneling occurs, this error is not very important. It  leads
only to a few percent error in calculation of the
temperature of the universe at the moment of the phase transition,
since the tunneling action is extremely sensitive
to even very small changes of the temperature. Thus, one may argue
that the thin wall approximation is still useful. (See also the talk
of Anderson at this Conference.) However, it is possible to determine
the time of the phase transition with an accuracy of few percent
without any study of tunneling: It is enough to say that the phase
transition happens  in the middle of the interval between $T_c$ and
$T_o$. In order to obtain a complete description of the phase
transition, including a correct shape of the bubble wall,
one should go beyond the thin wall approximation.

We would now like to obtain an analytic estimate of the probability
of tunneling in the electroweak theory,
which can be  used for any
particular numerical values of constants $D$, $E$ and
$\lambda_T$. As shown in Ref.  \cite{[5]},
eq. (\ref{7}) in most interesting cases approximates $V(\phi,T)$
with an accuracy of a few percent. This by itself does not help very
much if one must study tunneling  anew for each new set of the
constants. However, it proves possible to reduce this
study to the calculation of one function $f(\alpha)$, where $\alpha$
is some ratio of constants $D$, $E$ and $\lambda_T$. In what
follows we will calculate this
function for a wide range of values of $\alpha$. This will make it
possible to investigate tunneling in the electroweak theory without
any further use of computers.

First of all, let us represent the effective Lagrangian $L(\phi,T)$
near the point of the phase transition in the following form:
\begin{equation}\label{24}
L(\phi,T) =  {1\over 2} (\partial_{\mu}\phi)^2  -
{M^2(T)\over 2} \phi^2 + ET \phi^3 - {\lambda_o\over 4} \phi^4 \ .
\end{equation}
Here $M^2(T)  = 2 D (T^2 - T_o^2)$ is the effective mass squared of
the field $\phi$ near the point $\phi = 0$, $\la_t$ is the value
of the effective coupling constant $\lambda_T$ near the point of the
phase transition  (i.e. at $T \sim T_t$, where $T_t$ is the
temperature at the moment of tunneling). With a very good accuracy,
the constants
  $\lambda_t,  \lambda_{T_1}, \lambda_{T_c}, \lambda_{T_o}$ are equal
to each other.

Defining $\phi = {M^2\over 2 ET} \Phi$,
$x =  X/M$, the effective Lagrangian can be written as:
\begin{equation}\label{25} L(\Phi,T) = {M^6\over
4E^2T^2}\Bigl[{1\over 2} (\partial_{\mu} \Phi)^2  -
{1\over 2} \Phi^2 + {1\over 2} \Phi^3 - {\alpha\over 8} \Phi^4 \Bigr]
\ , \end{equation}
where
\begin{equation}\label{26}
\alpha = {\lambda_o M^2\over 2 E^2T^2}\ .
\end{equation}
The overall factor ${M^6\over 4E^2T^2}$ does not affect the Lagrange
equation
\begin{equation}\label{27}
{d^2\Phi\over dR^2} + {2\over R}\,{d\Phi\over dR} = \Phi -  {3\over
2}\Phi^2 +  {1\over 2}\alpha \Phi^3\ .
\end{equation}
Solving this equation and integrating over $d^3X = M^{-3} d^3x$
gives the following expression for the corresponding action:
\begin{equation}\label{28}
{S_3\over T} = {{4.85\,{M^3}}\over {{{ E}^2}\,{T^3}}} \times
f(\alpha) \ .
\end{equation}
The function $f(\alpha)$  is equal  \cite{[3]}
to 1 at  $\alpha= 0$, and blows up when $\alpha$ approaches 1.
In the whole interval from 0 to 1 this function, with an accuracy
about 2\%, is given by the following simple expression:
\begin{equation}\label{29}
f(\alpha) = 1 + {\alpha\over 4} \Bigl[ 1
+{{2.4}\over {1 - \alpha}} +
 {{0.26}\over{{{( 1 - \alpha ) }^2}}}\Bigr] \ .
 \end{equation}

In the vicinity of the critical temperature $T_o$, i.e. at  $\Delta T
\equiv T - T_o \ll T_o$, the action (\ref{28}) can be written in the
following form:
\begin{equation}\label{30}
 {S_3\over T} =  {38.8\,D^{3/2}\over {{{ E}^2}}}\cdot \left({\Delta
T\over T}\right)^{3/2} \times f\Bigl({2\,\lambda_o\,D\,\Delta T\over
E^2\,T}\Bigr) \ .
\end{equation}
Using these results, one can easily get analytical expressions for
the tunneling probability in a wide class of theories with
spontaneous symmetry breaking, including GUTs and the minimal
electroweak theory.

\section{Subcritical Bubbles}

Despite our semi-optimistic conclusions concerning the
infrared problem, it is still desirable to check that the whole
picture of the behavior of
the scalar field described above is (at least) self-consistent. This
means that
if the effective potential is actually given by eqs. (\ref{7}),
(\ref{8}),
(\ref{10}), (\ref{11}), (\ref{newE}), then our subsequent description
of the phase transition and
the bubble formation is correct. Indeed, one would expect that the
theory of
bubble formation is reliable, since the corresponding action for
tunneling
$S_3/T $ is very large, $S_3/T \sim 130 - 140$.  However, recently
even the validity of this basic assumption has been questioned.
Gleiser and Kolb \cite{[7]} and Tetradis  \cite{Tetr} have argued
that in many cases phase transitions occur not  due to bubbles of a
critical size, which we studied in section 3, but due to
smaller, subcritical bubbles.  We believe that these
authors raise a real issue.
However, we will now argue that this problem only arises if the
phase transition is extremely weakly first order.

The basic difference between the analysis of Ref.
\cite{[7],Tetr} and the more conventional one is
their assumption that at the time of the phase transition
there is a comparable probability to find different parts of the
universe in either of the two minima of $V(\phi,T)$. The main
argument of Ref. \cite{[7],Tetr} is that if the dispersion
of thermal fluctuations of the scalar
field $<\phi^2>\sim T^2$ is comparable with the distance between the
two minima of $V(\phi,T)$, then the field $\phi$ ``does not know''
which minimum is true and which is false. Therefore it spends
comparable time in each of them. According to  \cite{[7]},  a
kind of equilibrium between the
domains of the two types is achieved due to subcritical bubbles with
small action $S_3/T$ if many such bubbles may appear within a
horizon of a radius $H^{-1}$.

In order to investigate this question in a more detailed way, let us
re-examine our own assumptions concerning the
distribution of the scalar field $\phi$ prior to the moment at which
the temperature drops down to $T_1$, when the second minimum of
$V(\phi,T)$  appears. According to (\ref{13}), the value of the
scalar field
$\phi$ in the second minimum at the moment when it is formed is equal
to $\phi_1 = {3ET\over 2\lambda_T}$.  For $m_H \sim 60$ GeV
(and taking into account the coefficient $2/3$ in the cubic term) one
obtains $\phi_1 \sim 0.4\, T$. Thermal fluctuations of the
field $\phi$ have the dispersion squared $<\phi^2> = T^2/12$.
(Note an important factor $1/12$, which was absent in the
estimate made in \cite{[7]}.) This gives dispersion of thermal
fluctuations $\sqrt{<\phi^2>} \sim 0.3\, T$,
which is not much smaller than $\phi_1$.

However, as the authors of  \cite{[7]} emphasized in their
previous work  \cite{[12]} (see also  \cite{Tetr}), the total
dispersion $<\phi^2> \sim T^2/12$
is not an adequate quantity to consider since we are not really
interested in infinitesimally small domains containing different
values
of fluctuating field $\phi$. They argue that the proper measure of
thermal fluctuations is the contribution to $<\phi^2>$ from
fluctuations of the
size of the correlation length $\xi(T) \sim M^{-1}(T)$. This leads to
an
estimate $<\phi^2> \sim T\,M(T)$, which also may be quite large
 \cite{[12]}. Here again one should be very careful to use the proper
coefficients in the estimate. One needs to understand also why this
estimate could be relevant.

In order to make the arguments of Ref.  \cite{[7],Tetr} more
quantitative and to
outline the domain of their validity, it is helpful to review the
stochastic approach to tunneling (see  \cite{[10]} and
references
therein). This approach is not as precise as the euclidean approach
(in theories where the euclidean approach is applicable). However,
it is much simpler and more intuitive, and it may help us to look
from
a different point of view on the results we obtained in the previous
section and on the approach suggested in  \cite{[7],Tetr}.

The main idea of the stochastic approach  can be illustrated by an
example of tunneling with bubble formation from the point
$\phi = 0$ in the theory  (\ref{24})  with the effective potential
\begin{equation}\label{31}
V(\phi,T) = {M^2(T)\over 2} \phi^2 - ET\phi^3  +  {\lambda_o\over 4}
\phi^4 .
\end{equation}
For simplicity, we will study here the limiting case $\lambda_o\to
0$.

At the moment of its formation, the bubble wall does not move.
In the limit of small bubble velocity, the equation of motion of the
field $\phi$ at finite temperature is simply,
\begin{equation}\label{32}
\ddot\phi = d^2\phi/dr^2 + (2/r)d\phi/dr  - V^\prime(\phi) \  .
\end{equation}
The bubble starts growing if $\ddot\phi > 0$, which requires that
\begin{equation}\label{33}
 |d^2\phi/dr^2 + (2/r)d\phi/dr| < - V^\prime(\phi) \ .
\end{equation}
A bubble of a classical field is formed only if  it
contains a sufficiently big field $\phi$. It should be over the
barrier,  so that
$dV/d\phi < 0$, and the effective potential there should be
negative since otherwise formation of a bubble will be
energetically unfavorable. The last condition means
that the field $\phi$ inside the critical bubble should be
somewhat larger than  $\phi_*$, where $V(\phi_*,T) = V(0,T)$.
In the theory (\ref{31}) with $\la_o\to 0$, one has
$\phi_* = M^2/2ET$. As a simplest (but educated) guess,
let us take  $\phi \sim  2\phi_* = M^2/ET$.
Another important condition is that the size of the bubble
should be sufficiently large. If the size of the bubble
is too small, the gradient terms are bigger than the term
$|V^\prime(\phi)|$, and the field
$\phi$ inside the bubble does not grow. Typically, the second term in
 (\ref{33})
somewhat compensates the first one. To make a very rough estimate,
 one may write the condition (\ref{33}) in the form
\begin{equation}\label{34}
{1\over 2} r^{-2} \sim
{1\over 2} k^2 <  {1\over 2} k^2_{max}  \sim \phi^{-1}
|V^\prime(\phi)| \sim 2 M^2  .
\end{equation}
Let us estimate the probability of an event in which thermal
fluctuations
with $T \gg M$ build up a configuration of the field
satisfying this condition.
The dispersion of thermal fluctuations of the field
$\phi$ with $k<k_{max}$ is given by
\begin{eqnarray}\label{35}
<\phi^2>_{k<k_{max}} & = &
{1\over 2\pi^2}\int_{0}^{k_{max}}{k^2 dk\over
\sqrt {k^2 + M^2}\left(\exp{\sqrt{k^2 + M^2(\phi)}\over T} - 1
\right)} \nonumber \\
 & \sim & {T \over 2\pi^2}\int_{0}^{k_{max}}{k^2 dk\over
{k^2 + M^2}} \ .
\end{eqnarray}
Note that the main contribution to the integral is given
by $k^2 \sim k^2_{max} \sim 4 M^2 $.
This means that one can get a reasonably good estimate
of $<\phi^2>_{k<k_{max}}$ by omitting  $M^2$ in the
integrand. This also means that this estimate will be good
enough even though the effective mass  of the scalar field
$M^2(\phi)=V^{\prime\prime}(\phi)$ changes
between $\phi = 0$ and
$\phi$. The result we get is
\begin{equation}\label{36}
<\phi^2>_{k<k_{max}} \simeq    {T\over 2\pi^2}
\int_{0}^{k_{max}}{ dk}
  = { T k_{max}\over 2\pi^2} = {C^2   TM\over  \pi^2} \ .
\end{equation}
Here $C = O(1)$ is a coefficient reflecting the uncertainty
in the determination of  $k_{max}$ and estimating the integral.

Thus, we have a rough estimate of the dispersion of
perturbations which may  sum up
to produce a  field $\phi$ which satisfies the condition (\ref{34}).
We can  use it to evaluate the probability
that these fluctuations build up a bubble of the field $\phi$ of a
radius $r > k^{-1}_{max}$. This can be done with the help of the
Gaussian
distribution\footnote {The probability distribution is approximately
Gaussian even though the effective potential is not purely quadratic.
The reason is that we were able to neglect the curvature of the
effective potential {$m^2 = V^{\prime\prime}$} while calculating
{$<\phi^2>_{k<k_{max}}$.}}
\begin{equation}\label{37} P(\phi)
\sim \exp(-{\phi^2\over 2<\phi^2>_{k<k_{max}}}) =   \exp (-{
M^3\pi^2\over  2 C^2 E^2 T^3}) \sim \exp (-{4.92 M^3\over C^2
E^2 T^3}) \ .
\end{equation}
Note that the factor in the exponent in (\ref{37}) to within  a
factor of $C^2 = 1.02$  coincides with the exact result for  the
tunneling probability in this theory obtained by the euclidean
approach
 \cite{[3]} (see eq. (\ref{29})):
\begin{equation}\label{38}
P \sim \exp (-{4.85 M^3\over  E^2 T^3}) \ .
\end{equation}
Taking  into account  the very
rough method we used to calculate the dispersion of the perturbations
responsible for tunneling, the coincidence is rather impressive.

As was shown in \cite{[10]}, most of the results concerning tunneling
 at zero temperature,
at a finite temperature and even in the inflationary universe, which
were obtained by euclidean methods,
can easily be reproduced (with an accuracy  of the
coefficient  $C^2 = O(1)$ in the exponent) by this  simple method.

Now let us return to the issue of subcritical bubbles. As we have
seen, dispersion of the long-wave perturbations of the scalar field,
$<\phi^2>_{k<k_{max}} \simeq {k_{max}\,T\over 2\,\pi^2}$,
is quite relevant  to the theory of tunneling. Its calculation
provides a
simple and intuitive way to get the same results as we obtained
earlier by the euclidean approach  \cite{[10]}.  To get a good
estimate of the
probability of formation of a critical bubble in our simple model one
should calculate  this dispersion for $k_{max} \sim 2 M(T)$, which
gives $<\phi^2>_{k<k_{max}} = TM/\pi^2$. Note, that this
estimate is much
smaller than the naive estimate $<\phi^2> \sim TM$.

The crucial test of our basic assumptions  is a comparison
of this dispersion and the value of the field $\phi$ at
the moment $T = T_1$, when the minimum at $\phi   =
\phi_1 \not = 0$ first appears. Using eqs. (\ref{7}),
(\ref{13}), one can easily check that the mass of the
scalar field at $T = T_1$, $\phi = 0$ is given by
\begin{equation}
m = {3ET\over 2 \sqrt\lambda_T} \ .
\end{equation}
This yields
\begin{equation}
 \sqrt {<\phi^2>}_{k<k_{max}} \  \sim
 \phi_1\  {\lambda^{3/4}\over\pi\sqrt {3E/2}} \approx \phi_1 \
{10  \lambda_T^{3/4}\over  \pi } \ .
\end{equation}
For the Higgs boson with $m_H  \sim 60$ GeV one obtains
\begin{equation}
\sqrt {<\phi^2>}_{k<k_{max}} \ \sim   {\phi_1\over 5} \ .
\end{equation}
Thus, even with account taken of the factor $2/3$ in
the expression for $E$, the dispersion of long-wave fluctuations
of the scalar field is much smaller than the distance
between the two minima. Therefore,
the field $\phi$ on a scale equal to its correlation length
$\sim M^{-1}$ is not equally distributed between the two
minima of the
effective potential. It just fluctuates with a very small amplitude
near the point $\phi = 0$.
The fraction of the volume of the universe filled by the field
$\phi_1$ due to these fluctuations (i.e. due to subcritical bubbles)
for $m_H  \sim 60$ GeV is negligible,
\begin{equation}\label{ttt} P(\phi_1)
\sim\exp\left(-{\phi^2\over 2<\phi^2>_{k<k_{max}}}\right)
\sim\exp\left(-{3E\,\pi^2\over 4\lambda_T^{3/2}}\right)  \sim e^{-
12} \ {}.
\end{equation}
Since we already successfully applied this method for
investigation of tunneling, we expect that this estimate
is also reliable. The answer  remains rather small even
for $m_H \sim 100$ GeV, when
the phase transition is very weakly first order.

Moreover, even these long-wave fluctuations do not lead to
formation of stable
domains of space filled with the field $\phi \not = 0$,
until the temperature is below $T_c$ and
critical bubbles appear.
One expects a typical subcritical bubble to
collapse in a time $\tau \sim k_{max}^{-1}$; this is about
thirteen orders of magnitude smaller than
the total duration of the phase
transition, $\Delta t \sim 10^{-2} H^{-1} \sim 10^{-4} M_p \,
T^{-2}$.
We do not see any mechanism which might increase $\tau$ by
such a large factor.

Despite all these comments, we think that  subcritical
bubbles deserve further investigation.  They may lead to interesting
effects during phase transitions in GUTs, since the difference
between $T^{-1}$ and the duration of the GUT phase transitions
is not as great
as in the electroweak case. They may play an important role in  the
description of the electroweak phase transition as well, in models
where the phase transition occurs during a time not much
longer than $T^{-1}$. This may prove to be the case for very weakly
first order phase transitions with $10^3$ GeV $\gg \ m_H \, \,  \ga
\, \, 10^2$ GeV,
when the
distance between the two minima of $V(\phi,T)$ at $T \sim T_1$
is smaller than the dispersion
$\sqrt{<\phi^2>}_{k<k_{max}}\sim\sqrt{TM}/\pi$.

\section{Conclusions}
 One of the main consequences of our work \cite{OurPaper} is that it
is very difficult to generate baryon asymmetry in the  standard model
without expanding its Higgs sector. One the other hand, now we better
understand what is necessary for the electroweak baryogenesis to work
and how to calculate relevant quantities. Hopefully, this will help
us to find a realistic theory of elementary particles where
electroweak baryogenesis is possible.

\vskip 1cm

{\large \bf Acknowledgements}
This work could not be done without collaboration with Michael Dine,
Patrick Huet, Robert Leigh and Dmitri Linde. I appreciate very much
fruitful discussions with  Greg
Anderson, Renata Kallosh, Larry McLerran and Lenny Susskind.
I am grateful to Lawrence Krauss and Soo-Jong Rey for their
hospitality during the Conference in Yale.
This work was  supported in part by the National Science Foundation
grant PHY-8612280.


\vskip 2cm

\begin{thebibliography}{999}

\bibitem{OurPaper} M. Dine, R. Leigh, P. Huet, A. Linde and D. Linde,
Stanford University preprints
 SU-ITP-92-6 (1992) (to be published in Physics Letters)
 and SU-ITP-92-7 (1992) (to be published in Phys. Rev.).
\bibitem {[1]} D.A. Kirzhnits, JETP Lett. {\bf 15} (1972) 529;
 D.A. Kirzhnits and  A.D. Linde, Phys. Lett. {\bf 42B} (1972) 471.
\bibitem {[1a]} S. Weinberg, Phys. Rev. {\bf D9} (1974) 3357;
 L. Dolan and R. Jackiw, Phys. Rev. {\bf D9} (1974) 3320;
 D.A. Kirzhnits and  A.D. Linde, JETP {\bf 40} (1974) 628.
\bibitem {[1b]} D.A. Kirzhnits and  A.D. Linde, Ann. Phys. {\bf 101}
 (1976) 195.
\bibitem{Linde81} A.D. Linde,  Phys.Lett. {\bf 99B} (1981) 391.
\bibitem {[11]} A.D. Linde, Rep. Prog. Phys. {\bf 42} (1979) 389.
\bibitem{Guth81} A.H. Guth, Phys. Rev. {\bf D23} (1981) 347; \\
 A.D. Linde, Phys. Lett. {\bf 108B} (1982); {\bf 114B} (1982) 431;
 {\bf 116B} (1982) 335, 340; \\ A. Albrecht and P.J. Steinhardt,
 Phys. Rev. Lett. {\bf 48} (1982) 1220.
\bibitem {[2]} A.D. Linde, {\em Particle Physics and Inflationary
 Cosmology} (Harwood, Chur, Switzerland, 1990).
 \bibitem{Tunn1} A.D. Linde, Phys.Lett. {\bf 70B}  (1977) 306.
 \bibitem{DS} S. Dimopoulos and L. Susskind, Phys. Rev.
 {\bf D18} (1978) 4500.
 \bibitem{KRS} V.A. Kuzmin, V.A. Rubakov and M.E. Shaposhnikov,
 Phys. Lett. {\bf B155} (1985) 36; P. Arnold and  L. McLerran, Phys.
Rev. {\bf D36}{581}{87}.
\bibitem{5} M.E. Shaposhnikov, JETP Lett. {\bf 44} (1986) 465;
 Nucl. Phys. {\bf B287} (1987) 757;  Nucl. Phys. {\bf B299} (1988)
797;
 A.I. Bochkarev, S.Yu. Khlebnikov and M.E. Shaposhnikov,
 Nucl. Phys. {\bf B329} (1990) 490.
\bibitem{6} L. McLerran, Phys. Rev. Lett. {\bf 62} (1989) 1075.
\bibitem{7} L. McLerran, M. Shaposhnikov, N. Turok and
 M. Voloshin, Phys. Lett. {\bf 256B} (1991) 451.
\bibitem{8} N. Turok and P. Zadrozny, Phys. Rev. Lett.
 {\bf 65} (1990) 2331; Nucl. Phys. {\bf B358} (1991) 471.
\bibitem{9} M. Dine, P. Huet, R. Singleton and L. Susskind,
 Phys.Lett. {\bf 257B} (1991) 351.
\bibitem{10} A. Cohen, D.B. Kaplan and A.E. Nelson, Nucl. Phys.
 {\bf B349} (1991) 727.
\bibitem{11} A. Cohen, D.B. Kaplan and A.E. Nelson, Phys.Lett.
 {\bf 263B} (1991) 86.
\bibitem{12} A. Cohen, D.B. Kaplan and A.E. Nelson, University of
 California, San Diego, preprint  UCSD-PTH-91-20 (1991)
\bibitem{[8]} A. Bochkarev, S. Kuzmin and M. Shaposhnikov, Phys.
 Lett. {\bf 244B} (1990) 27.
\bibitem{LEP} ALEPH, DELPHI, L3 and OPAL Collaborations,
 as presented by M. Davier, Proceedings of the International
 Lepton-Photon Symposium and Europhysics Conference on
 High Energy Physics, eds. S. Hegerty, K. Potter and E. Quercigh
 (Geneva, 1991), to appear.
\bibitem {[4]} M. Dine, P. Huet and R. Singleton,   Nucl. Phys. {\bf
B375} (1992) 625; A.D.  Linde and D.A. Linde, unpublished.
\bibitem{[5]} G. Anderson and L. Hall, Phys. Rev. {\bf D45} (1992)
625.
\bibitem{Hsu} D. Brahm and S. Hsu, Caltech preprints CALT-68-1705 and
 CALT-68-1762 (1991).
\bibitem{Shap} M.E. Shaposhnikov, Phys. Lett {\bf B277} (1992) 324.
\bibitem {Carr} M.E. Carrington, Phys. Rev. {\bf D45} (1992) 2933.
\bibitem {[3]} A.D. Linde, Phys.Lett. {\bf 70B}  (1977) 306;  {\bf
 100B} (1981) 37; Nucl. Phys. {\bf B216} (1983) 421.
\bibitem {[7]} M. Gleiser and E. Kolb,  preprint
 FERMILAB-Pub-91/305-A (1991).
\bibitem{Tetr} N. Tetradis, preprint DESY 91-151.
\bibitem{Kaj} K. Enqvist, J. Ignatius, K. Kajantie, K. Rummukainen,
Phys.Rev. {\bf D45} (1992) 3415.
\bibitem {sher} M. Sher, Phys. Rep. {\bf 179} (1989) 273.
\bibitem {[6]} S. Coleman, Phys. Rev. {\bf D15} (1977) 2929.
\bibitem{L} A.D. Linde, Phys. Lett. {\bf 93B} (1980) 327.
\bibitem{LGPY}
 D.J. Gross, R.D. Pisarski and L.G. Yaffe, Rev. Mod. Phys. {\bf 53}
 (1981) 1.
\bibitem{EQZ} J.R. Espinosa, M. Quiros and F. Zwirner,
 preprint CERN-TH.6451/92 (1992).
 \bibitem{BHG} G. Boyd, D.E. Brahm and D.H. Hsu, preprint
CALT-68-1795 (1992).
\bibitem{evans} T.S. Evans, Imperial/TP/91-92/23 (Apr.\ 1992).
\bibitem {[12]} M. Gleiser, E. Kolb and R. Watkins,
 Nucl. Phys. {\bf B364} (1991) 411.
\bibitem {[10]} A.D. Linde,  Nucl. Phys. {\bf B372} (1992) 421.


\end{thebibliography}
\end{document}